\documentclass[aps,pre,nofootinbib,superscriptaddress,nopacs,amsmath,amssymb,final,letterpaper,notitlepage]{revtex4-1}

\usepackage[utf8]{inputenc}
\usepackage{calc}
\usepackage{graphicx}
\usepackage{amsmath,amssymb,amsthm}
\usepackage{bbold}
\usepackage{bm}
\usepackage{color}
\usepackage[dvipsnames]{xcolor}
\usepackage{enumitem}
\usepackage{tikz}
\usepackage{float}
\usepackage[percent]{overpic}
\usepackage{parskip}
\usepackage{adjustbox}
\usepackage{multirow}
\usepackage{appendix}
\usepackage{booktabs} 
\usepackage{caption}  
\usepackage{siunitx}  
\usepackage{geometry}

\makeatletter
\renewcommand{\section}{%
  \@startsection{section}{2}{0pt}%
  {\baselineskip}
  {0.5\baselineskip}
  {\normalfont\normalsize\bfseries\raggedright}
}
\makeatother

\makeatletter
\renewcommand{\subsection}{%
  \@startsection{subsection}{2}{0pt}%
  {\baselineskip}
  {0.5\baselineskip}
  {\normalfont\normalsize\bfseries\raggedright}
}
\makeatother

\makeatletter
\renewcommand{\subsubsection}{%
  \@startsection{subsubsection}{2}{0pt}%
  {\baselineskip}
  {0.5\baselineskip}
  {\normalfont\normalsize\bfseries\raggedright}
}
\makeatother

\let\oldabstract\abstract
\let\endoldabstract\endabstract
\renewenvironment{abstract}
  {\oldabstract\vspace{-\baselineskip}} 
  {\endoldabstract\vspace{-\baselineskip}} 

\usepackage{enumitem}

\def\multiset#1#2{\ensuremath{\left(\kern-.3em\left(\genfrac{}{}{0pt}{}{#1}{#2}\right)\kern-.3em\right)}}

\usepackage[colorlinks=true,linkcolor=blue,urlcolor=blue,citecolor=blue,anchorcolor=blue]{hyperref}

\begin{document}
\title{Where is AIED Headed? Key Topics and Emerging Frontiers (2020-2024)}

\author{Shihui Feng}
\email{shihuife@hku.hk}
\affiliation{Faculty of Education, University of Hong Kong, Pok Fu Lam Road, Hong Kong, China}
\affiliation{Institute of Data Science, University of Hong Kong, Pok Fu Lam Road, Hong Kong, China}

\author{Huilin Zhang}
\affiliation{Faculty of Education, University of Hong Kong, Pok Fu Lam Road, Hong Kong, China}

\author{Dragan Gašević}
\affiliation{Faculty of Information Technology, Monash University, Clayton, VIC, Australia}

\maketitle

{\bfseries}\quad
In this study, we analyze 2,398 research articles published between 2020 and 2024 across eight core venues related to the field of Artificial Intelligence in Education (AIED). Using a three-step knowledge co-occurrence network analysis, we analyze the knowledge structure of the field, the evolving knowledge clusters, and the emerging frontiers. Our findings reveal that AIED research remains strongly technically focused, with sustained themes such as intelligent tutoring systems, learning analytics, and natural language processing, alongside rising interest in large language models (LLMs) and generative artificial intelligence (GenAI). By tracking the bridging keywords over the past five years, we identify \textbf{four emerging frontiers in AIED—LLMs, GenAI, multimodal learning analytics, and human-AI collaboration}. The current research interests in GenAI are centered around GAI-driven personalization, self-regulated learning, feedback, assessment, motivation, and ethics.The key research interests and emerging frontiers in AIED reflect a growing emphasis on  \textbf{co-adaptive, human-centered AI for education}. This study provides the first large-scale field-level mapping of AIED’s transformation in the GenAI era and sheds light on the future research development and educational practices. 


\vspace{7pt}

\section{Introduction}

The field of Artificial Intelligence in Education (AIED) has undergone rapid development in recent years, driven by advances in machine learning, natural language processing, and the increasing integration of AI into educational platforms. AIED as an interdisciplinary research field transcends disciplinary boundaries and integrates methods and tools from computer science and information science to solve educational problems (Pinkwart, 2016; Feng \& Law, 2021; Hwang et al., 2020). Its applications span diverse educational scenarios—from formal classroom instruction to informal self-directed learning—enabling dynamic, personalized knowledge dissemination across contexts while supporting process-oriented assessment and real-time feedback (Guan et al., 2024; Ayemowa et al., 2024; Dai et al., 2024; Hwang et al., 2020).\\

The release of GPT-3 by OpenAI in 2020 marked a significant progress in generative artificial intelligence (GenAI), showcasing unparalleled capabilities in language understanding and generation (Aryadoust et al., 2024). This advancement made the generation of human-like content feasible, leading to widespread applications across various fields, including education (Watson \& Shi, 2024). The advent of large language models (LLMs) has expanded AI applications in educational contexts beyond traditional tasks such as automated grading and intelligent question answering to advanced functions like generating personalized lesson plans, delivering real-time learning recommendations, and creating immersive experiences for language acquisition (Alaqlobi et al., 2024; Salih et al., 2024; Wang et al., 2025). These advancements are redefining the research paradigms and practical applications of AIED, paving the way for new opportunities and possibilities in the future of education. \\

Given the changes happening in the field, tracking shifts in research focus of AIED—through the lens of keyword co-occurrence networks—offers valuable insights into emerging priorities, educational applications, and future research directions. Feng \& Law (2021) analyzed the keyword co-occurrence network of 1,830 articles in the AIED field from 2010 to 2019. They identified that the sustained themes in AIED were intelligent tutoring systems and massive open online courses, and highlighted the emerging topics including neural networks, deep learning, eye tracking, and personalized learning during that period. In retrospect, the development of neural networks and deep learning has enabled the success of GenAI, which provides foundation models trained on large-scale dataset and propels AI applications in education to grow rapidly in recent years. Built on this previous work, in the current study, we analyzed the keyword co-occurrence network of 2,398 articles from eight key venues related to AIED for the period 2020-2024, to identify the recent key research focus and uncover the intrinsic connections among diverse interests within the field. Specifically, this study was guided by the following three research questions: 

\begin{itemize}
    \item[]\textbf{RQ1:} What is the knowledge structure of AIED between 2020 and 2024 (five years)?
    \item[]\textbf{RQ2:}What are the key knowledge clusters of AIED between 2020 and 2024?
    \item[]\textbf{RQ3:} What are the emerging frontiers in AIED?
\end{itemize}

\section{Related Work}
With the ongoing expansion of AIED research, comprehensive reviews are needed for elucidating the key topics in the field and offering directions for future studies. Several review articles have summarized major research topics in the field of AIED. For instance, Zhai et al. (2021) conducted a content analysis of 100 studies published between 2010 and 2020, identifying major research dimensions such as classification, recommendation, deep learning, real-time feedback, and adaptive learning. They also outlined four critical future research directions: the Internet of Things, collective intelligence, deep learning, and neuroscience. Mustafa et al. (2024) reviewed 143 reviews in the AIED field, using the technology-enhanced learning model to code the literature. Their findings indicate that AI applications in education primarily target higher education, teacher support, and student learning, while receiving less attention for administrators, school leaders, and special education. \\

In addition, there are reviews focusing on specific tools or application scenarios within AIED, including evaluating the features and applications of AI tools such as intelligent tutoring systems (ITS), automated scoring systems, recommendation systems, and learning analytics. Mousavinasab et al. (2018) examined 53 papers published between 2007 and 2017, systematically summarizing variations and technological foundations of ITS developed for diverse educational contexts. Huawei and Aryadoust (2023) analyzed 105 articles published before May 2020, assessing the application contexts, effectiveness, and limitations of automated scoring systems. Additionally, Additionally, some studies employed meta-analyses to evaluate the impact of AI tools, revealing significant effects on learning outcomes, medium improvements in higher-order thinking (HOT) skills, and smaller but notable influences on learning perceptions (Alemdag, 2023; Dibek et al., 2024; Zheng et al., 2021).\\

The application scenarios of AIED are also broad, encompassing medical education, STEM education, language teaching, K-12 education, mathematics education, and higher education (Hwang \& Tu, 2021; Xu \& Ouyang, 2022; Zafari et al., 2022).  For example, some researchers have used narrative or scoping reviews to explore the forms of AI applications, student experiences, and limitations in medical education (Preiksaitis \& Rose, 2023). Xu and Ouyang (2022) conducted a systematic review of 63 empirical AI-STEM studies from 2011 to 2021, summarizing six categories of AI applications and illustrating the relationships between AI application types and other elements (e.g., information, subjects, media, and environments) within AI-STEM systems. Similarly, Zafari et al. (2022) performed a systematic review of 210 articles and conference papers indexed in the Web of Science and Scopus databases from 2011 to 2021. They categorized AI applications in K-12 education into four dimensions: student performance, teaching, selection and behavior tasks, and others.\\

A few studies employed a bibliometric approach to analyze AIED literature, aiming to have a comprehensive description of the research field. For instance, Delen et al. (2024) analyzed 4,673 publications from 1975 to 2023 involving 29 AIED-related keywords, providing a multidimensional visualization of annual publication patterns, disciplinary distribution, and leading authorship within the field. Responding to the recent rise of GenAI, Polat et al. (2024) conducted a bibliometric study of 212 Scopus-indexed publications up to July 2023, identified via the keywords “ChatGPT” and “education.” The study revealed ChatGPT’s exponential growth in education with focuses on  computational intelligence, human learning, language processing, and conversational agents. Bozkurt et al. (2021) applied social network analysis and text-mining techniques to 276 research articles, identifying three key research clusters: artificial intelligence, pedagogical issues, and technological issues. They also proposed five major research themes, including “adaptive learning and personalization of education through AI-based practices.” Feng and Law (2021) analyzed 1,830 articles published in five core venues of AIED for the period 2010-2019 and employed a three-level keyword co-occurrence network analysis to reveal the knowledge structure and examine the evolution of knowledge clusters in two-year intervals, offering data-driven insights into mapping the AIED landscape.\\

Despite these existing reviews in the field, there are gaps that this study aims to address. First, existing studies predominantly rely on keyword-based searches in databases, with the number of keywords ranging from a few to several dozen, varying greatly across studies (eg. Delen et al., 2024; Xu \& Ouyang., 2022). Given that AIED is a highly interdisciplinary field with a diverse array of keywords, relying on keyword-based article selection often fails to comprehensively capture the full scope of AIED research. Second, there is a lack of comprehensive review mapping the recent and updated AIED development with large scale data. Given the rapid evolution of AIED over the past five years—particularly with the rise of GenAI—there is a need to reassess the field’s current development and emerging trends. Therefore, this study aimed to address these gaps, and map the recent research focus and emerging topics in the field of AIED. 

\section{Methods}
\subsection{Data Collection and Pre-processing}
In this study, we collected all publications from 2020 to 2024 from eight core publication venues related to AIED, including (1) International Conference on Artificial Intelligence in Education (AIEDC); (2) International Journal of Artificial Intelligence in Education (IJAIED); (3) International Conference on Educational Data Mining (EDM); (4) ACM Conference on Learning at Scale (L@S); (5) International Conference on Intelligent Tutoring Systems (ITS); (6) Journal of Learning Analytics (JLA); (7) International Learning Analytics \& Knowledge Conference (LAK) proceedings; (8) Computer and Education: Artificial Intelligence (C\&EAI). The selection of these eight venues was determined through consultations with leading experts in the field. These eight publication sources comprehensively span the intersection of education and artificial intelligence, with coverage ranging from theoretical exploration to technological innovation and real-world application, aiming to provide a holistic representation of developments in AIED research.\\

We collected all articles published in these eight sources between 2020 and 2024 to analyze research topics and trends in AIED. Detailed information of each source is provided below:
\begin{itemize}
    \item AIEDC: 484 full and short papers
    \item IJAIED:  256 articles
  \item EDM: 280 full and short papers (not including posters)
  \item L@S: 363 articles
  \item ITS: 269 articles
  \item JLA: 157 articles
  \item LAK: 379 full and short papers 
  \item C\&EAI: 331 articles
\end{itemize}

Since this study maps recent AIED developments through keyword network analysis, we performed rigorous data pre-processing on the retrieved article keywords. We first excluded the articles containing more than 10 keywords or lacking keywords from the dataset. After that, we pre-processed the keywords to ensure consistency and eliminate redundancy based on the following steps:

\begin{enumerate}
\item Converting to lowercase and removing hyphens
\item Converting plurals into a singular form—Using the WordNetLemmatizer from Python’s Natural Language Toolkit (NLTK), plural keywords were transformed into their base forms. While this process achieves high accuracy, rare errors such as improperly transforming "SES" into "S" still existed and were resolved through manual verification.
\item Replacing abbreviations with their corresponding full forms—We used two approaches to replace the abbreviations: (a) extracting keywords with parentheses (e.g., "explainable ai (xai)") and removing their abbreviations, and (b) manually verifying keywords with fewer than four characters or written in uppercase to identify abbreviation candidates.
\item Merging synonyms—Keywords with identical meanings but different expressions were unified to ensure consistency in representing research content. Synonymous keywords were identified using Python’s rapidfuzz library with a similarity threshold of 90, followed by manual validation. Four types of synonym variations were addressed, including (a) different spellings (e.g., “human centred computing” vs. “human centered computing”), (b) typographical errors (e.g., “principle component analysis” vs. “principal component analysis”), (c) compound words and split words (e.g., “clickstream” vs. “click stream”), and (d) semantically equivalent expressions (e.g., “automated assessment” vs. “automatic assessment”). \\
\end{enumerate}

After the data pre-processing, there are 4,733 distinct keywords from the 2,398 articles in the final dataset for the following analysis.

\subsection{Data Analysis}
We employed the three-step approach of keyword co-occurrence network (KCN) analysis proposed by Feng and Law (2021) to address the research questions in this study. The three step-approach analyzes a constructed KCN at the macro-, meso- and micro-levels, to uncover the knowledge structure, knowledge clusters and emerging topics of a reviewed research field.  A Keyword Co-occurrence Network (KCN) is a weighted graph representation of keyword relationships in academic literature, where nodes correspond to keywords and edges reflect co-occurrence frequencies. The network topology of a KCN reveals the knowledge structure of a field, while node-level metrics quantify keyword importance (Su \& Lee, 2010; Radhakrishnan et al., 2017). KCNs are considered an effective tool for large-scale knowledge mapping within a research field (Radhakrishnan et al., 2017). \\

To address RQ1, we analyzed the structural characteristics of the constructed KCN. First, we examined the distribution of weighted degrees against a power law distribution to assess whether the network follows a scale-free topology. Second, we investigated the relationship between node degree—the number of connections a keyword (node) had—and its weighted local clustering coefficient—a measure of how densely interconnected a node’s neighbors are, weighted by the strength of their associations. This analysis helped determine whether the network exhibited a structure where a small subset of keywords had many sparsely connected associations, while the majority of keywords had fewer but more densely connected associations. Lastly, we investigated the tendency of nodes to connect with others of higher or lower degrees by analyzing the relationship between node degrees and their average weighted nearest neighbor degree. We employed a modified average weighted nearest neighbor degree measure (Feng \& Law, 2021), which calculated the average degree of a node’s neighbors while accounting for edge weights. The modified version further divided this value by the focal node’s degree to assess homophily tendencies at the node level.\\

To address RQ2, we applied modularity-based clustering to identify community structures within the KCNs, revealing cohesive research themes in AIED. Communities (or clusters) are network subgroups where nodes exhibit denser internal connections than external ones, indicating shared characteristics (Fortunato, 2010). In KCNs, keywords clustered together represented distinct knowledge domains within AIED. We implemented the fast-greedy algorithm (Clauset et al., 2004) on each KCN’s largest connected component to optimize computational efficiency and reduce noise. This approach automatically determined the optimal number of communities by maximizing modularity—a metric quantifying partition quality by comparing intra- vs. inter-cluster edge density (Newman, 2006). Higher modularity values indicated more meaningful community divisions. The fast-greedy algorithm efficiently handled large networks via hierarchical agglomeration while optimizing global modularity (Clauset et al., 2004), making it ideal for mapping structural shifts in AIED’s research landscape over time. Each identified knowledge cluster was named after the keyword with the highest in-group degree. At the meso-level, we analyzed the static knowledge clusters of AIED between 2020 and 2024, as well as the temporal knowledge clusters of the KCNs for each year to traced knowledge evolution over time.\\

To address RQ3, we analyzed the temporal KCNs for each year at the micro-level to identify trending and bridging keywords using weighted betweenness centrality. Keywords exhibiting high weighted betweenness centrality act as gatekeepers, bridging disparate keyword clusters and maintaining knowledge connectivity across AIED's research landscape. The nodes showing sudden increases in betweenness centrality were considered new trending topics (Yang et al., 2014). \\

Furthermore, to provide a deeper understanding of the themes and clusters identified through the KCN analysis, we incorporated content analysis as a complementary qualitative approach. For each knowledge cluster obtained from modularity-based clustering, we closely read and summarized the articles containing the cluster’s core keywords, synthesizing their research objectives, methodologies and key findings to elucidate the substantive content within each cluster. For the emerging keywords with high betweenness centrality, we also retrieved and carefully reviewed literature from the entire database containing these keywords, focusing on their application scenarios, methodological innovations, and primary challenges. This approach enabled us to enrich our interpretation of the application features and potential impacts of these concepts within the AIED field, and to outline possible directions for their future development.

\section{Results}
\subsection{Most Frequent Keywords in AIED Field (2020-2024)}
The top 20 keywords with the highest frequency in AIED from 2020 to 2024 (the last five years) are shown in Figure 1. Keywords including machine learning, natural language processing (NLP), LLMs, GenAI, and deep learning feature prominently, underscoring the technical focus of AIED research. Regarding educational implications, the period saw strong interest in concepts including self-regulated learning, assessment, feedback, collaborative learning, and online learning. During the last five years of 2020-2024, AI rapidly rose to become the second most frequent keyword. The emergence and rising prominence of keywords such as LLM, GPT, and GenAI reflect the recent advances in foundational AI technologies and their rapid adoption in educational settings.\\

\begin{figure*}
    \centering
    \includegraphics[width=0.9\textwidth]{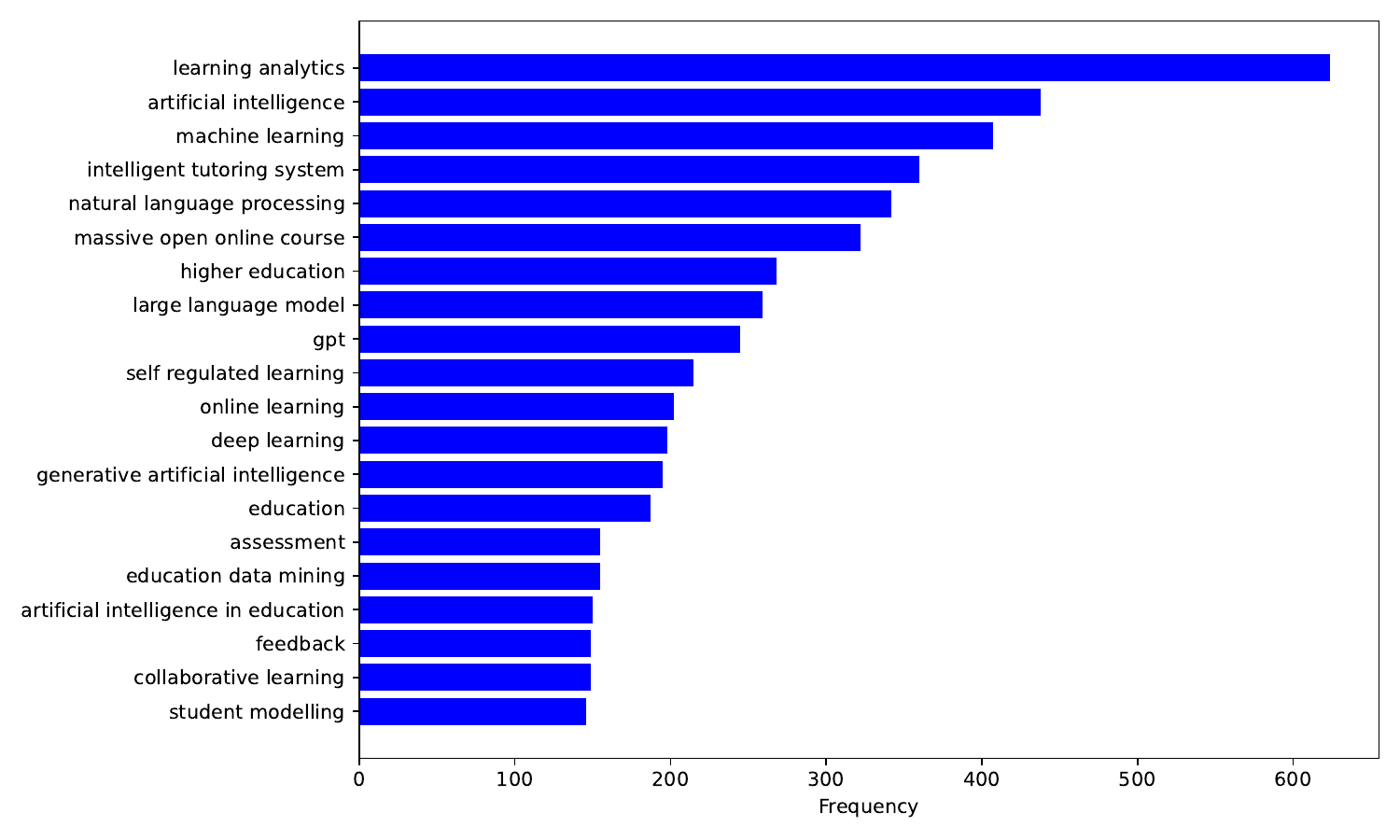}
    \caption{Top 20 keywords with the highest frequency
    }
    \label{fig:diagram}
\end{figure*}

\subsection{Knowledge structures of the AIED field (2020-2024) (RQ1) }

The structural characteristics of the constructed KCN reveal the underlying knowledge structure of the AIED field. Our analysis showed that the KCN exhibited a hierarchical structure, characterized by two key properties: a heavy-tailed degree distribution and a high clustering coefficient (Ravasz \& Barabási, 2003). This structure suggests that a small subset of keywords dominated the literature, while the majority appear only sporadically.\\

We summarize the structural features of the temporal and overall KCNs in Table 1. The network metrics provide insights into the evolving connectivity and thematic organization of AIED research. The yearly KCNs consistently showed low density (sparse connections) but high local clustering (local keyword groupings). The average degree (z) and average strength (s) have showedn a fluctuating upward trend over the past five years, reaching their peak in 2024. This indicates that both the breadth and the intensity of connections among research topics have further increased in recent years. The largest component (lc) refers to the maximal subset of nodes in a network in which every pair of nodes can be reached via paths. The high lc values approaching the total number of nodes (n) in each year indicates strong interconnectedness among . It is an important indicator for assessing the degree of interconnectedness among topics keywords within a fieldwithin AIED.  Overall, in the past five years, the proportion of the largest component to the total number of nodes has reached 95\%, with annual proportions ranging from 82.9\% to 93.1\%. This demonstrates that there weare extensive connections and interactions among research topics within the field of AIED. The degree correlation coefficient (r) of the network wais negative, indicating a certain level of disassortativity; that is, nodes with high degrees tended to connect with those with low degrees. This pattern likely ariseds because mainstream topics (eg. ‘natural language processing’, ‘artificial intelligence’)  served as “hubs” within the network, frequently co-occurring with a range of specialized or emerging topics.\\

\begin{table}[htbp]
  \centering
  \caption{Structural Characteristics of the Keyword Co-Occurrence Networks}
  \label{tab:network_characteristics}
  \begin{tabular*}{\textwidth}{@{\extracolsep{\fill}} l 
    S[table-format=4.0]   
    S[table-format=6.0,group-digits=false]   
    S[table-format=1.3]   
    S[table-format=1.3]   
    S[table-format=1.3]   
    S[table-format=1.3]   
    S[table-format=4.0]   
    S[table-format=-1.3]  
  @{}}
    \hline
    \textbf{Years} & \textbf{\emph{n}} & \textbf{\emph{m}} & \textbf{\emph{d}} & \textbf{\emph{c}} & \textbf{\emph{z}} & \textbf{\emph{s}} & \textbf{\emph{lc}} & \textbf{\emph{r}} \\
    \midrule
    \hline
    2020 & 1270 & 3519 & 0.004 & 0.819 & 5.542 & 6.173 & 1085 & -0.055 \\
    2021 & 1350 & 3803 & 0.004 & 0.824 & 5.634 & 6.424 & 1119 & -0.068 \\
    2022 & 1222 & 3272 & 0.004 & 0.803 & 5.355 & 5.975 & 1036 & -0.088 \\
    2023 & 1244 & 3483 & 0.005 & 0.809 & 5.600 & 6.625 & 1095 & -0.076 \\
    2024 & 1560 & 5061 & 0.004 & 0.819 & 6.489 & 7.551 & 1452 & -0.120 \\
    All & 4733 & 191138 & 0.002 & 0.841 & 8.087 & 9.262 & 4497 & -0.099 \\
    \hline
  \end{tabular*}
  
  \vspace{0.5em}
  \small
  \textit{Note.} $n$ = number of nodes; $m$ = number of edges; $d$ = network density; $c$ = average clustering coefficient; \\
  $z$ = average degree; $s$ = average weighted degree; $lc$ = the size of largest component; $r$ = degree person correlation coefficient (average assortativity).
\end{table}

The weighted degree distribution of the KCN exhibited a power-law distribution, as shown in Figure 2. The vast majority of nodes had low weighted degrees and were connected to only a few other nodes, while only a small number of hub nodes possessed extremely high weighted degrees and accounted for a large proportion of the network's connections. In this study, the fitted power-law exponent was 2.28. As noted by Newman (2003), power-law exponents for networks typically fall within the interval between 2 and 3, and the result of the overall KCN was consistent with this empirical regularity.\\

\begin{figure*}
    \centering
    \includegraphics[width=0.7\textwidth]{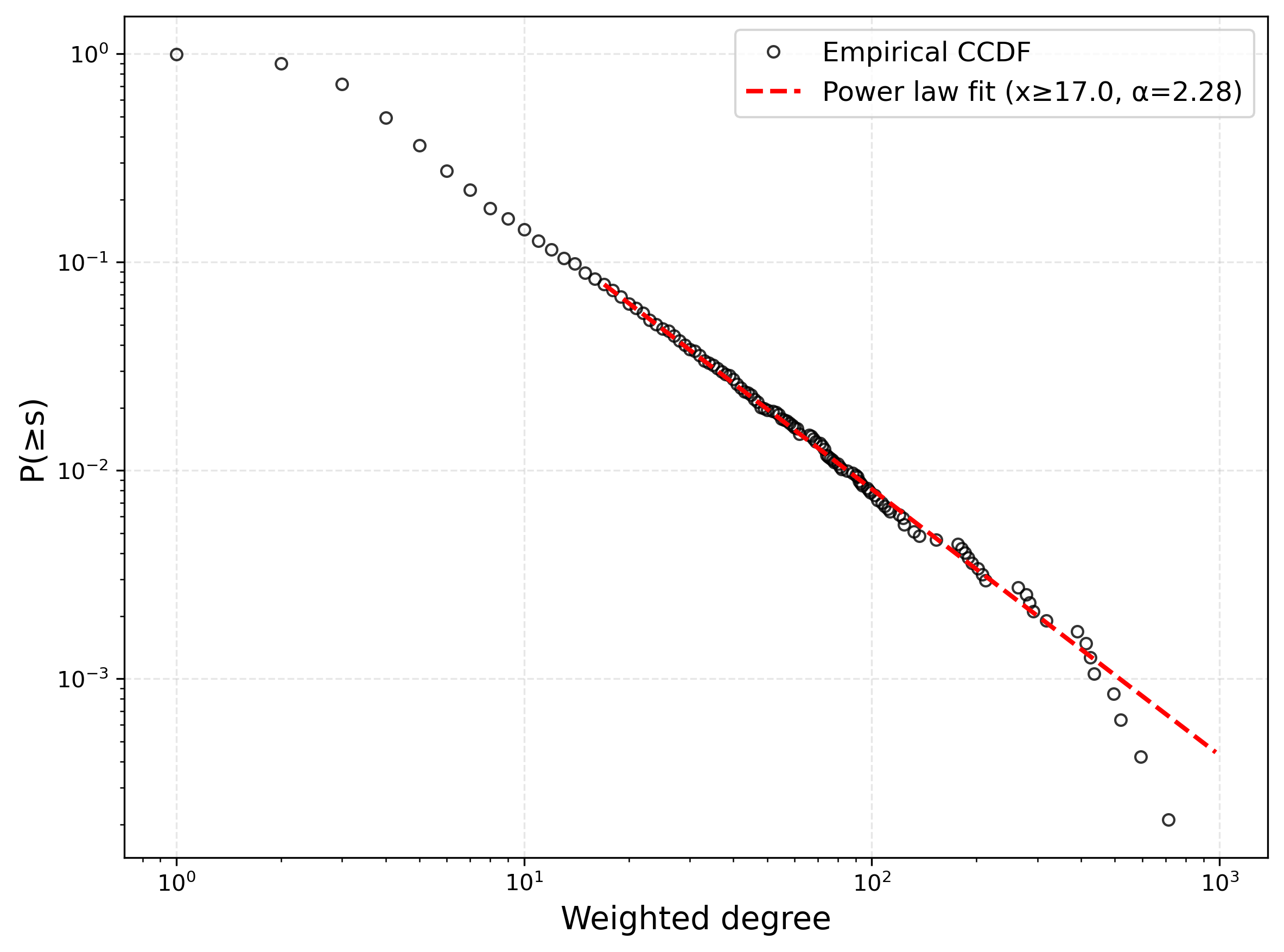}
    \caption{Complementary cumulative distribution function of weighted degrees (circles), with corresponding power law fit (dashed red line).
    }
    \label{fig:diagram}
\end{figure*}

Figure 3 illustrates the relationship between node degree and the weighted clustering coefficient in the KCN. As node degree increased, the weighted clustering coefficient declined markedly, reflecting a significant negative correlation between these two metrics. This trend suggests that high-degree nodes, although connected to numerous neighbors, were characterized by a relatively low likelihood of interconnections among their neighbors, resulting in sparser neighborhood structures and weaker local clustering. Conversely, the limited neighbors of low-degree nodes were more likely to form densely interconnected groups, thus exhibiting higher weighted clustering coefficients and indicating more cohesive local structures.\\

\begin{figure*}
    \centering
    \includegraphics[width=0.7\textwidth]{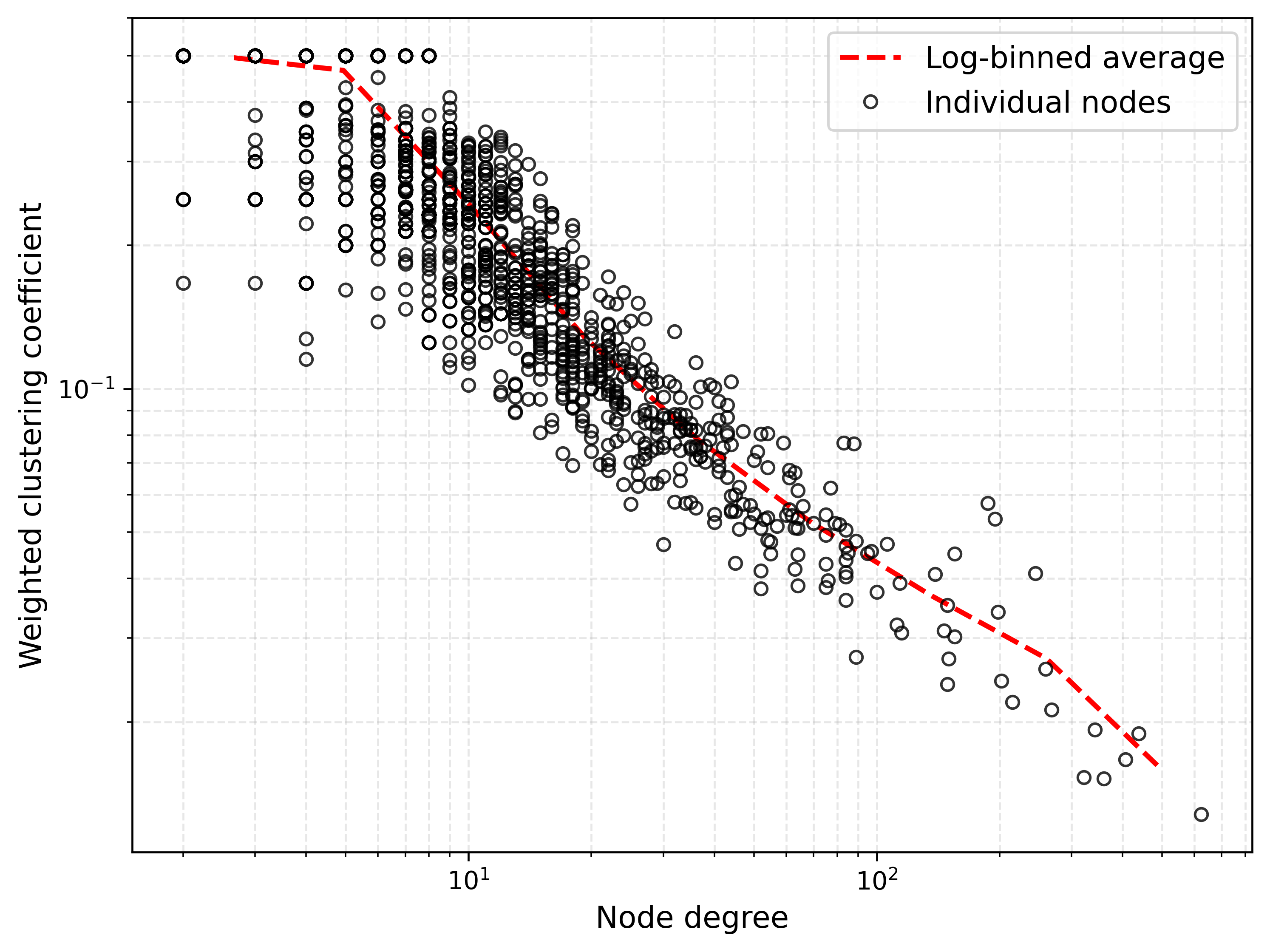}
    \caption{Node weighted clustering coefficient vs. node degree (circles), with corresponding average for nodes with similar degree (dashed red line).
    }
    \label{fig:diagram}
\end{figure*}

The relationship between the weighted average nearest neighbor degree and node degree is shown in Figure 4. The red dashed line represents the reference line where the ratio equals 1, serving as a threshold to distinguish the association tendencies of nodes. It can be seen that most nodes in the network tended to be connected with nodes of higher degree. As the node degree increased, the ratio of the weighted average nearest neighbor degree to the node’s own degree generally showed a decreasing trend. This indicates that nodes with higher degrees were more likely to connect with nodes of lower degree.\\

\begin{figure*}
    \centering
    \includegraphics[width=0.7\textwidth]{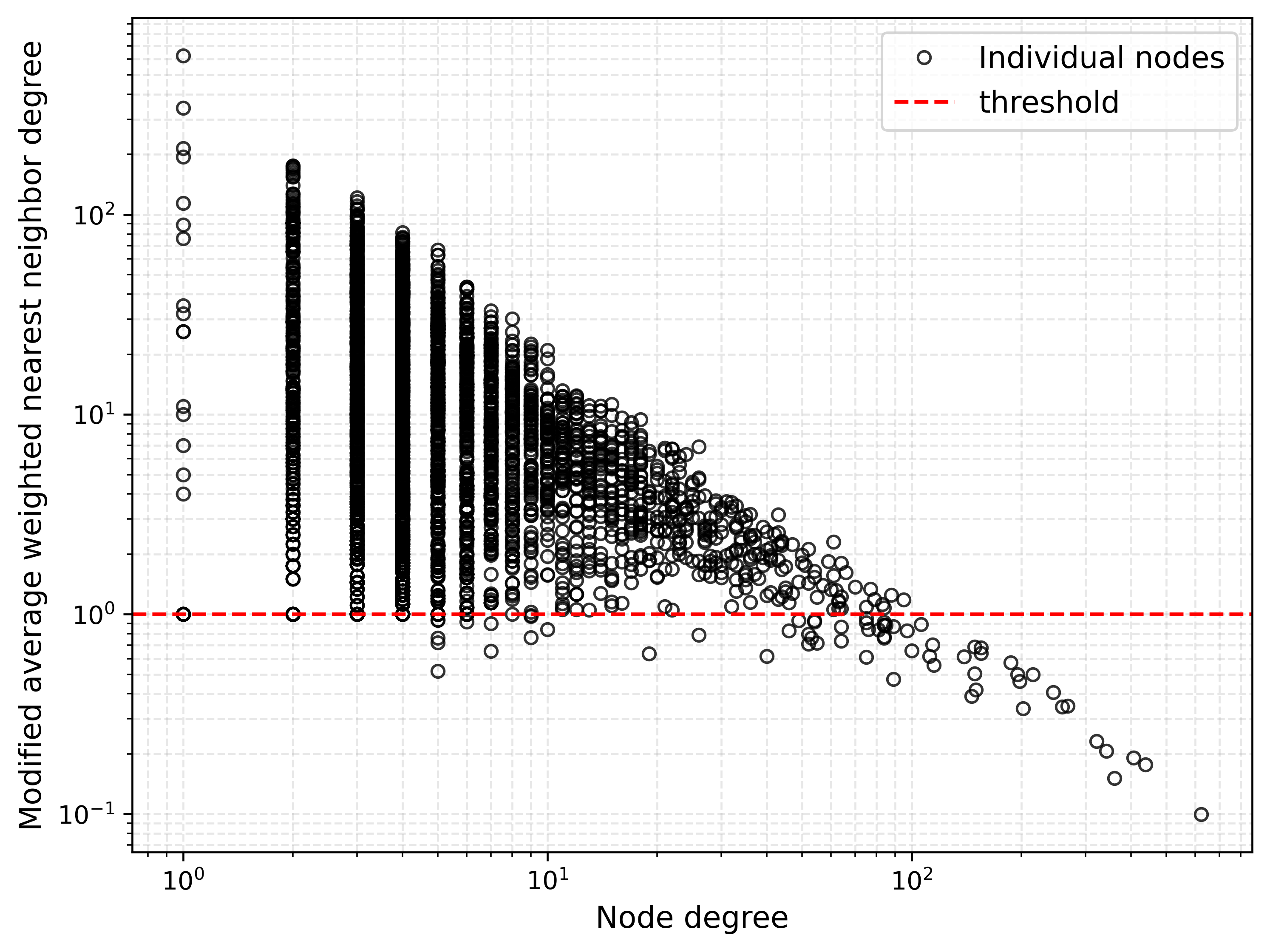}
    \caption{Average weighted nearest neighbor degree vs. node degree (circles), with a threshold dashline indicating the tendency to associate with lower or higher degree nodes.
    }
    \label{fig:diagram}
\end{figure*}

\subsection{Knowledge Clusters of the AIED Field (2020-2024) (RQ2)}

Frequency analysis alone cannot capture the relationships among keywords. We therefore constructed the KCN with the 4,733 distinct keywords from the 2,398 articles published between 2020 and 2024 and conducted the meso-level analysis to identify the knowledge clusters of the AIED field. We also constructed a temporal KCN for each year, and analysed the knowledge clusters of each temporal network to present the developmental trajectory of the field. \\

Figure 5 below presents the network visualization of the knowledge clusters of AIED research from 2020 to 2024, with the top ten largest knowledge clusters labelled in the graph. The most prominent knowledge clusters over the last five years were ‘natural language processing’, ‘‘learning analytics’, massive open online courses’, ‘artificial intelligence’, ‘engagement’, ‘intelligent tutorial system’, ‘causal inference’, ‘lifelong learning’, ‘computer vision’, and ‘data science application in education’. A table in the Appendix shows the top 10 highest in-group degree keywords for each knowledge cluster. These keywords represent the most central and frequently co-occurring terms within each cluster and support the interpretation of each knowledge cluster.\\

\begin{figure*}
    \centering
    \includegraphics[width=0.8\textwidth]{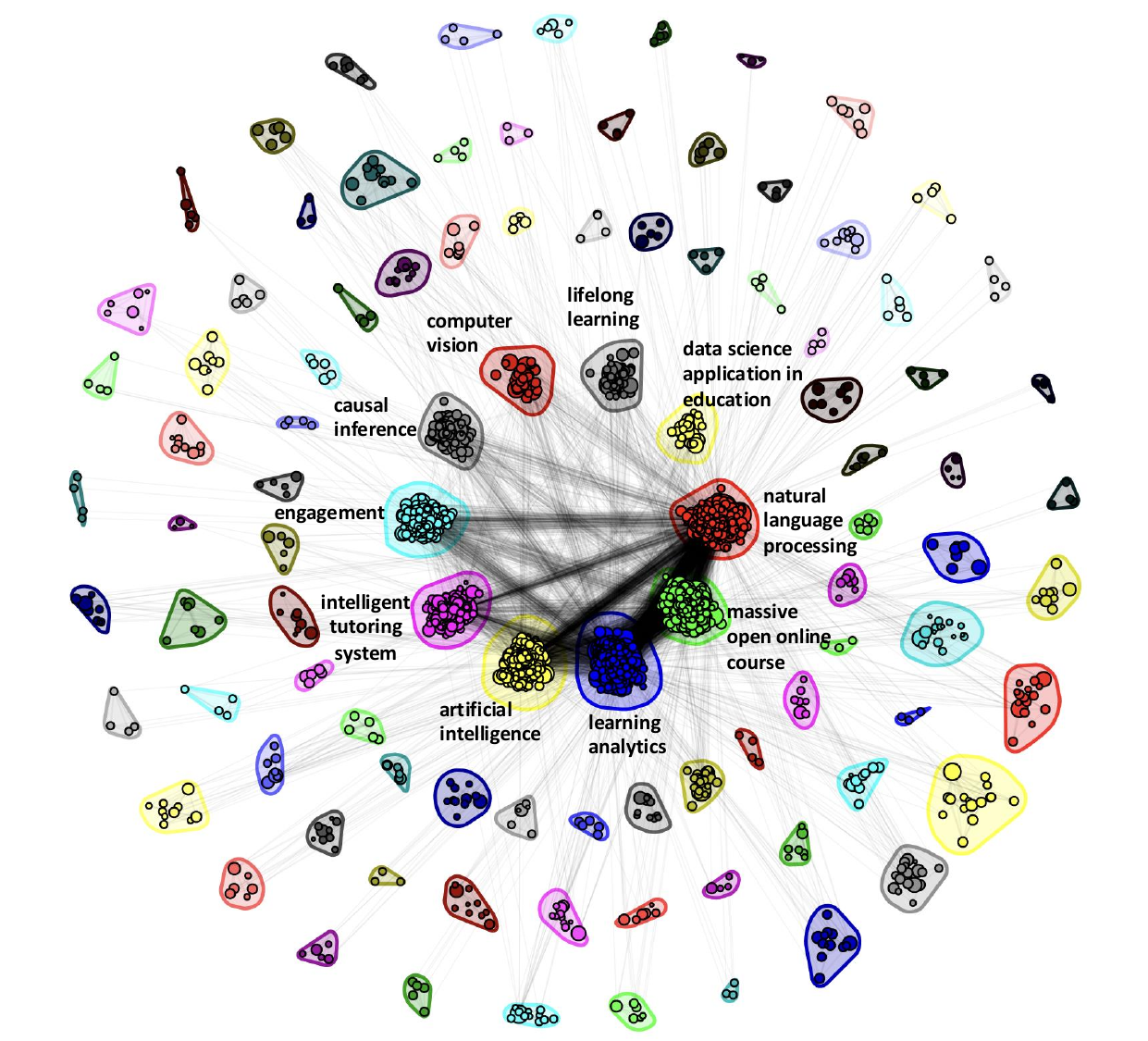}
    \caption{Knowledge clusters of AIED field from 2020-2024, with the top 10 largest knowledge clusters labelled in the graph.
    }
    \label{fig:diagram}
\end{figure*}

It is noteworthy that Figure 5 shows numerous interconnections of varying thickness among the knowledge clusters, reflecting both knowledge exchange and interdisciplinary integration within the field. The thicker links between clusters such as ‘natural language processing’, ‘learning analytics’, ‘artificial intelligence’, ‘massive open online courses’, and ‘intelligent tutoring systems’ suggest that these clusters frequently intersected and jointly drove innovation. For instance, intelligent tutoring systems were increasingly leveraging NLP and AI technologies for automated essay scoring and feedback, conversational Q\&A, and the detection of knowledge mastery during learning (Albano et al., 2023; Smerdon, 2024; Vasić et al., 2023). Likewise, learning analytics was widely used on MOOC platforms to track and analyze learners’ paths, interaction logs, and post content, enabling the prediction of student performance and personalized recommendations (Hsu et al., 2022; Thomas et al., 2022).  \\

To examine trends in knowledge clusters, the current study also produced chronological visualizations of the temporal knowledge cluster networks from 2020 to 2024 (see Figure 6 below). Comparison of these visualizations reveals that topics such as ‘natural language processing’, ‘learning analytics’, ‘massive open online courses’, ‘intelligent tutoring systems’ and ‘self regulated learning’ remained consistently prominent areas of focus, underscoring the sustained interests in these topics within the AIED field. \\

\begin{figure*}
    \centering
    \includegraphics[width=1.0\textwidth]{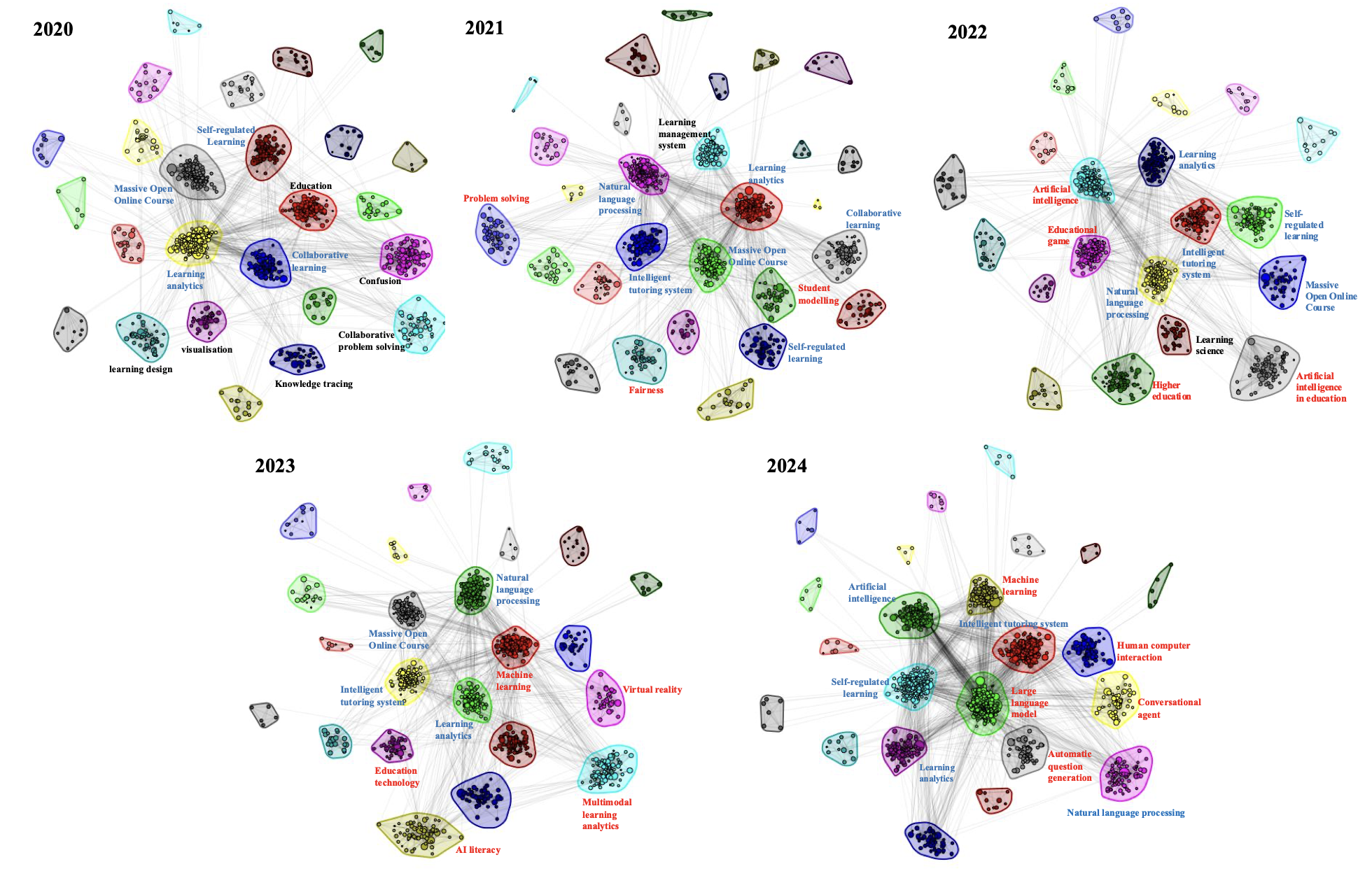}
    \caption{Temporal knowledge clusters of AIED field from 2020-2024
    }
    \label{fig:diagram}
\end{figure*}

 \textbf{‘Natural language processing’(NLP)} was the largest knowledge cluster in AIED research from 2020 to 2024. Building on advancements in machine learning and neural networks, NLP reached new heights with the launch of BERT and GPT. BERT, with its powerful bidirectional context understanding and fine-tuning capabilities, was particularly well-suited for tasks that require deep comprehension and analysis, such as text classification, and sentiment analysis (Cochran et al., 2023; Funayama et al., 2023; Ndukwe et al., 2020; Rodrigues et al., 2024). \\

\textbf{‘Learning analytics’ (LA)} enhanced the understanding of the learning process by collecting single-mode or multimodal data through educational data mining, while also providing robust support for student modeling. Learning analytics was widely applied in diverse contexts to identify patterns related to self-regulated learning, collaborative learning, and feedback provision (Boulahmel et al., 2024; Chandrasekaran et al., 2022; Jin et al., 2024).\\

\textbf{‘Massive open online course’ (MOOC)} is a significant form of online learning and has been extensively implemented in higher education contexts. Researchers utilize MOOC platforms to collect detailed data on student behaviors, which enables the construction of personalized learner profiles and the provision of tailored learning support (Valle Torre et al., 2020; Meaney \& Fikes, 2022; Ahmed et al., 2024b). Furthermore, ethical considerations remain a significant concern. Scholars investigated various strategies to safeguard student privacy, including local data storage and the implementation of Differential Privacy techniques to protect sensitive information (Valle Torre et al., 2020; Salman \& Alexandron, 2024).\\

 \textbf{‘Intelligent tutoring system’ (ITS)} remained a core and enduring cluster in the field.  Many intelligent tutoring systems are capable of structuring complex knowledge domains—such as aviation, and programming—using ontological approaches as a basis for assessing student performance or making content recommendations (Feitosa et al., 2024; Litovkin et al., 2022; Tchio et al., 2024). Some studies developed cognitive agents based on the ACT-R cognitive architecture integrated with an ontological reference model to simulate how humans perceive, understand, and perform tasks in authentic settings (Hayashi \& Shimojo, 2022; Tchio et al., 2024). By integrating intelligent algorithms, these systems can simulate, guide, automate, and recommend instructional activities, enabling more advanced instructional and cognitive support (Chanaa et al., 2020; Courtemanche et al., 2023).\\

\textbf{’Self-regulated learning’ (SRL)} appeared to be a key knowledge cluster within the field. It refers to the process by which students periodically adjust their learning behaviors to achieve specific learning goals, and its application in online education is particularly widespread. Some researchers integrated SRL with learning analytics LA to examine students' behavioral patterns during this process using multi-source data (Ng et al., 2023; Quick et al., 2023). Other studies explored methods to enhance SRL skills, such as incorporating learner confidence feedback and learner control to facilitate adaptive practice (Yan et al., 2024a), utilizing technology-supported instructional strategies like videos and prompts, and integrating chatbots (Lin \& Chang, 2023). Additionally, some researchers investigated the factors that influence SRL, including personality traits (Weng et al., 2024) and reflective abilities (Li et al., 2023).\\

Based on the analysis of knowledge clusters for the temporal KCNs, we noticed that several new knowledge clusters gained a growing interest within the field, including large language model, conversational agent, automatic question generation, human computer interaction, multimodel learning analytics, virtual reality, and AI literacy.\\

\textbf{‘Large language model’ (LLM)} represents an emerging knowledge cluster in 2024, highlighting the field’s fast adaptation to technical development. The 2023 release of high-performance, permissively licensed open-source models such as LLaMA 1 and LLaMA 2 was rapidly embraced by educational researchers to fine-tune these models for subjects like language learning, mathematics and science, establishing them as transformative tools in educational contexts (Lee et al., 2024; Watson \& Shi, 2024). \\

\textbf{‘Conversational agent’ (CA)} interacts with learners through text or verbal communication for a given designed learning pru. Many studies have explored various design approaches, such as enhancing learner control (through learning paths and interface appearance), altering agent language and text structures, integrating augmented reality to simulate real-world scenarios, and developing dialogue agents based on learning analytics and fuzzy rules, to examine their effects on learners' cognitive and non-cognitive development (Du \& Daniel, 2024; Li et al., 2024; Sosnowski et al., 2023; Wambsganss et al., 2024). CA technology evolved from its initial reliance on pattern-matching based on regular expressions to complex systems that leverage machine learning for enhanced user input recognition (Sosnowski et al., 2023). Notably, the rise of GenAI breathed new life into CA. Kakar et al. (2024) proposed a ChatGPT-based CA that demonstrates improved answer accuracy and reduced false positive rates. Additionally, they introduced methods to mitigate hallucinations and harmful content.\\

\textbf{‘Automatic question generation’ (AQG)} produce formative practice questions tailored to learning texts, aligning with the “learning by doing” philosophy thus effectively enhancing learning outcomes (Johnson et al., 2024a). By leveraging various NLP techniques, such as the TextRank algorithm, combined with templates, a rich variety of question types—including fill-in-the-blank, multiple-choice, and matching questions— can be generated (Brown et al., 2024; van Campenhout et al., 2023). With the advancement of AI technologies, researchers have begun exploring the use of GPT to assist in question generation and provide feedback (Diwan et al., 2023; Van Campenhout et al., 2024). Throughout this process, particular attention has been paid to the quality and appropriateness of the generated questions (Johnson et al., 2024b).
‘Human-Computer Interaction’ (HCI) primarily focuses on how human users effectively communicate with computer systems within educational contexts. At the core of HCI were users, with an emphasis on their needs, capabilities, and behaviors. For instance, Wang et al. (2023) highlighted that the monotony of existing sign language learning systems hinders effective communication between hearing individuals and the deaf community, leading them to develop an interactive system that enhances engagement in learning. Their research implemented diverse interaction methods, such as gesture recognition (Wang et al., 2023) and capturing human facial movements (Nishida et al., 2022), while also utilizing environmental simulators to improve user experience (Bonyad Khalaj et al., 2024). Additionally, other studies explored the impact of different interface designs on the interaction experience (Sheel et al., 2024). \\

\textbf{‘Multimodal learning analytics’ (MMLA)} emerged as a new knowledge cluster independent from learning analytics in 2023, underscoring the importance of analyzing and modeling multi-source, large-scale, and complex educational data. Enabled by advancements in sensor technology, computer vision, and artificial intelligence, learning analytics transcended the limitations of relying solely on a single data source consisting of digital traces (such as clicks, time spent, and scores), leading to a shift towards finer-grained and more dynamic learning analytics (Ouhaichi et al., 2023). This has led to its widespread application in various contexts, including online learning, collaborative learning, and game-based learning. \\

\textbf{‘Virtual reality’ (VR)}technology offers unique advantages for education through its high level of immersion. By creating virtual learning environments characters that interact with users, VR can enhance the acquisition of professional knowledge and improve the overall learning experience (Ng et al., 2022; Seo et al., 2023). Chen et al. (2023) specifically focused on commercial VR educational games and pointed out the deficiencies in effective feedback mechanisms in existing VR games. VR creates task load scenarios that can more realistically elicit participants' psychological states. Some researchers have explored the effectiveness of using technologies such as electroencephalography (EEG) and eye-tracking to detect learner distraction and mental fatigue in VR environments, as well as investigating corresponding interventions (Assaf et al., 2023; Zarour et al., 2023). These efforts provided valuable technological support for attention management in intelligent tutoring systems.\\

\textbf{‘AI literacy’} is critical for equipping students with the skills necessary to navigate future societal transformations. Southworth et al. (2023) introduced the UF AI Literacy Model, including enabling AI, know \& understand AI, use \& apply AI, evaluate \& create AI, and AI ethics. Due to the limitations of K-12 learners' foundational knowledge and cognitive abilities, researchers have developed varied definitions and curricular designs for AI literacy (Su et al., 2023; Yim, 2024a; Yim, 2024b). Some researchers have also developed various tools to assess students' AI literacy (Hornberger et al., 2023; Zhang et al., 2024a).

\subsection{Emerging Frontiers in the AIED Field (RQ3)}

Newly emerged keywords with high betweenness centrality merit special attention, as they reveal nascent interdisciplinary topics that bridge distinct fields and foster new research synergies. In this study, by tracing the top 20 highest weighted betweenness centrality nodes across the last five years of 2020 to 2024, we identified the emerging frontier of the AIED field based on the keywords that first appeared in the top 20 list (see Figure 7 below). The four emerging frontiers were—\textbf{large language model, generative artificial intelligence, human-AI collaboration, and multimodal learning analytics}. These were particularly important for signalling future directions in artificial intelligence in education. We extracted the ego network of each emerging themes and label the top 20 alters with the highest degrees to support the interpretation of the emerging themes. The reminder of the section provides a detailed analysis of these emerging topics.\\

\begin{figure*}
    \centering
    \includegraphics[width=0.8\textwidth]{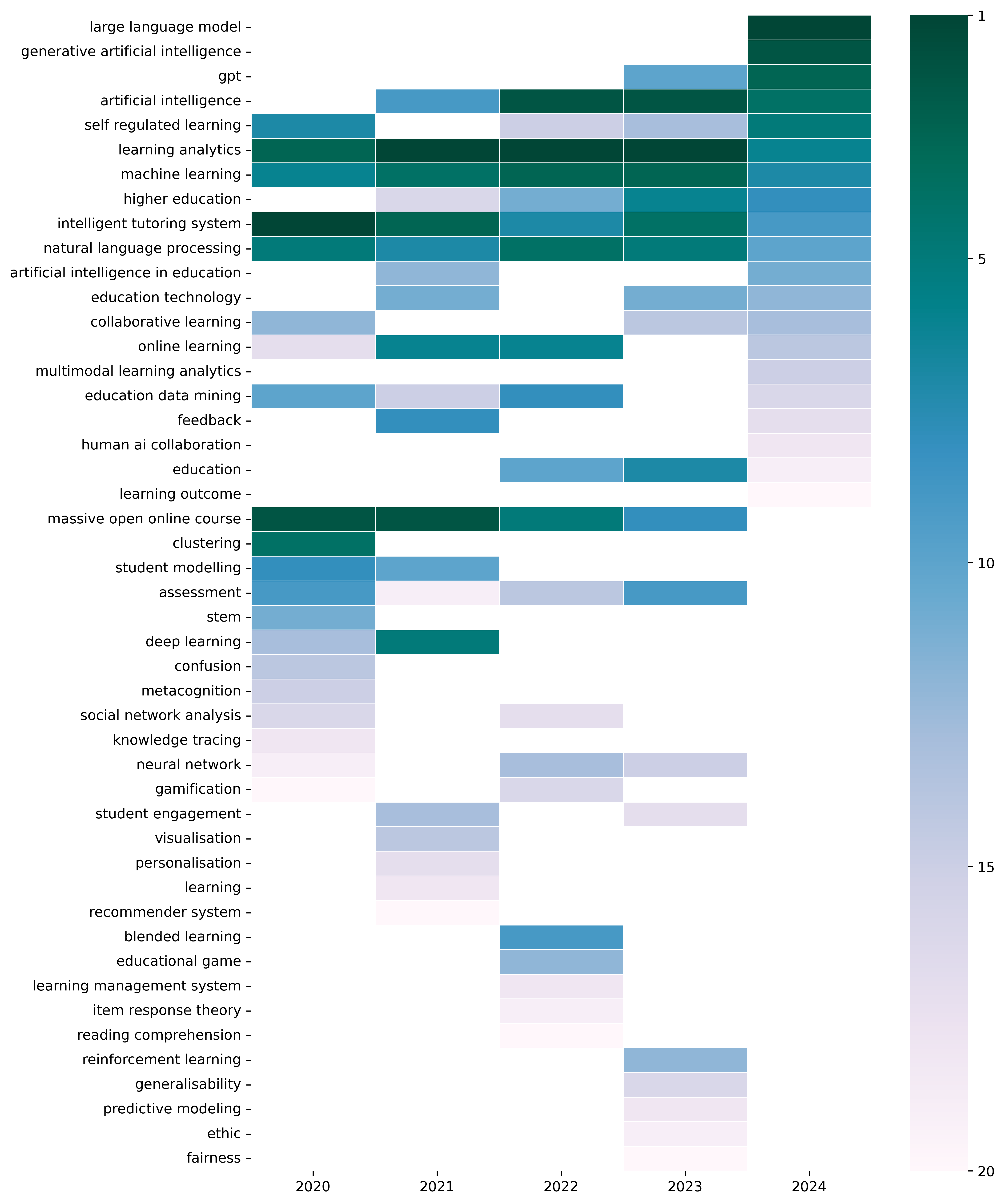}
    \caption{Top 20 highest weighted betweenness centrality keywords across five time periods
    }
    \label{fig:diagram}
\end{figure*}

\begin{figure*}
    \centering
    \includegraphics[width=0.7\textwidth]{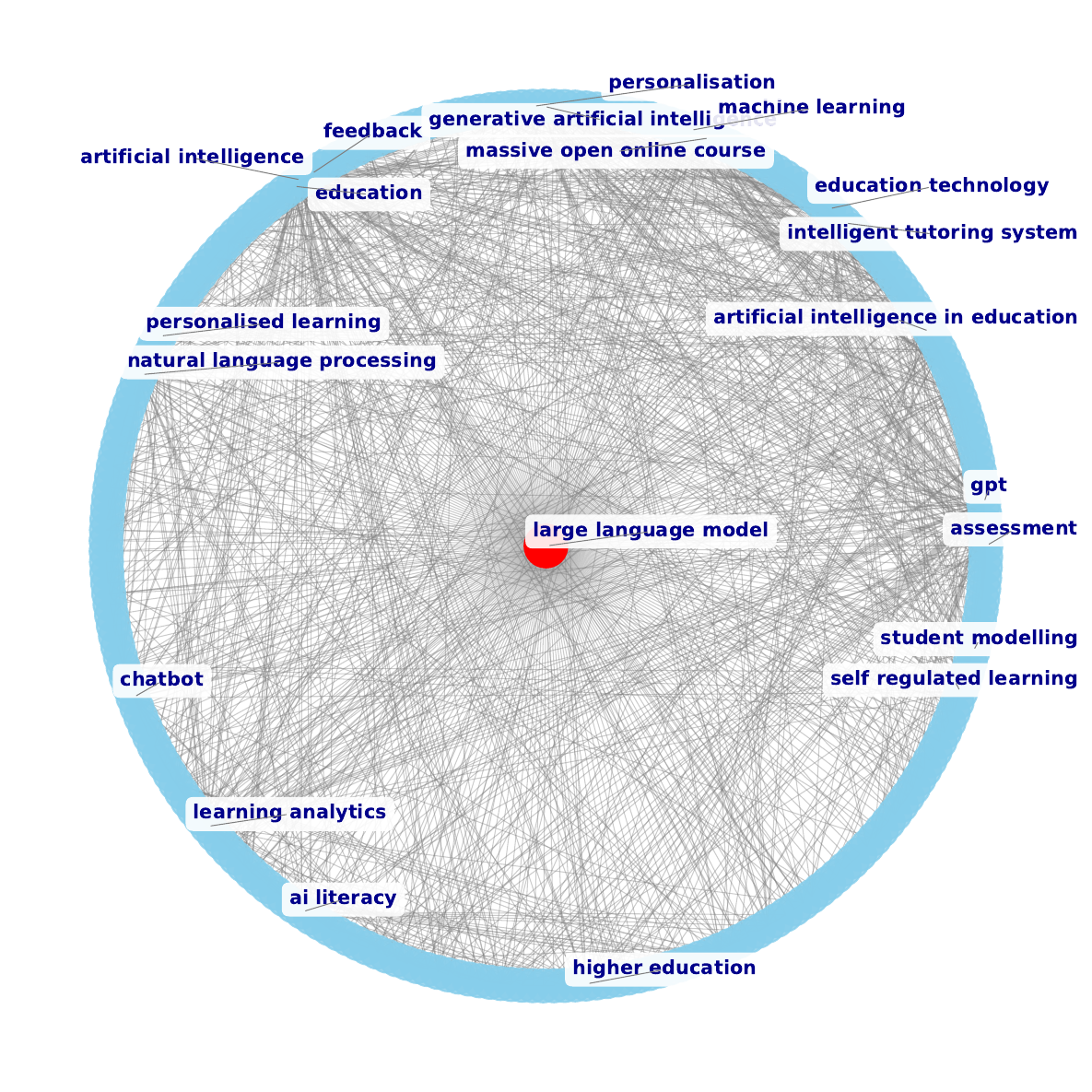}
    \caption{The ego network of the first emerging theme \textit{large language model}, with labels for the top 20 alters by degrees.
    }
    \label{fig:diagram}
\end{figure*}

The first emerging frontier was \textbf{large language models (LLMs)} (see Figure 8). It became the keyword with the highest betweenness centrality in 2024 for the first time, reflecting its rising prominence in AIED research. LLMs are built using machine learning—especially deep learning techniques—and serve as a cornerstone for modern NLP, such as sophisticated language understanding and generation. LLMs have been integrated with educational chatbots and are widely applied in areas such as instruction and feedback, greatly improving the quality and adaptability of their responses. For instance, Wang et al. (2024) proposed a learning framework with a chatbot-based integrated development environment forprogramming and testing that uses LLMs and strategy games to generate real-time training data and data-driven improvements. \\

The integration of LLMs with traditional AIED research domains, such as intelligent tutoring systems and MOOCs, is becoming increasingly prominent. For instance, a ChatGPT-based intelligent tutoring system, PyTutor, employs a phased strategy—including pseudocode, fill-in-the-blank exercises, and tiered code hints—to simulate teacher guidance and adapt to varying levels of task difficulty and student proficiency (Yang et al., 2024). This approach has been shown to significantly enhance students' classroom engagement and assignment completion rates (Yang et al., 2024). In addition to simulating the role of a teacher, LLMs also have the potential to act as learning partners, enriching the learning experience by offering diverse perspectives on problem-solving (Arnau-Blasco et al., 2024).  Providing learners with timely and effective feedback remains a significant challenge in large-scale online courses. To address this issue, automated grading functionalities based on LLMs are increasingly being integrated into online learning platforms to enhance the intelligence and efficiency of assignment evaluation. The multi-agent architecture EvaAI proposed by Lagakis and Demetriadis (2024), as well as the Gipy application for programming assessment developed by Gabbay and Cohen (2024), have both explored this area in depth. The analysis of LLMs applications across educational domains revealed its considerable potential for promoting personalized learning. Unlike traditional personalized learning systems, which require complex design, LLMs substantially lowers the barriers to personalized learning technologies. LLMs leverage powerful pre-trained models and flexible conversational interfaces to assess learning states and dynamically generate contextually relevant dialogues, prompts, thereby fostering dynamic, immersive learning experiences (Johnson, 2024; Li et al., 2024b; Naik et al., 2024). LLMs can thoroughly analyze students' background information and historical problem-solving data to achieve more accurate student modeling (Nguyen et al., 2024). Beyond cognition, it can also incorporate emotion recognition to offer tailored encouragement and support motivation (Gaeta et al., 2024).  \\

The second emerging frontier in the field is \textbf{Generative Artificial Intelligence} (see Figure 9) which is a significant branch of artificial intelligence research. From the ego network of GenAI, we found that the field’s current research interests of GenAI are centered around GenAI-driven personalization, self-regulated learning, feedback, assessment, motivation, and ethics. Large language models like GPT-3 exemplify this area, alongside other foundation models for vision and audio (Gupta et al., 2024; Paaß \& Giesselbach, 2023). As GenAI encompasses LLMs, certain alters in the ego networks overlap, such as higher education, assessment, feedback, learning analytics and self-regulated learning.  Due to the flexibility and self-regulation inherent in university-level instruction, higher education serves as a prime setting for experimenting with GenAI (Xiaoyu et al., 2025). Most studies primarily examine policies, current applications, challenges, and future directions, with key challenges including faculty competence, hardware infrastructure, and equity (Chiu, 2023; Jin et al., 2024b; Wang et al., 2024a). Given that GenAI is a relatively new technology, researchers are especially interested in the motivation of both teachers and students to engage with it.  Studies have found that both groups generally acknowledge the supportive role of GenAI in learning, and that intrinsic motivation (such as interest and a sense of satisfaction) as well as external expectations and self-concept can enhance their willingness to use these tools (Lai et al., 2023; Huang \& Mizumoto, 2024; Nikou et al., 2024). However, evidence regarding whether GenAI can increase students’ motivation to engage with course content remains inconclusive, underscoring the need for further development of motivation measurement instruments tailored to AI-related contexts  (Guan et al., 2024; Lai et al., 2023). \\

\begin{figure*}
    \centering
    \includegraphics[width=0.7\textwidth]{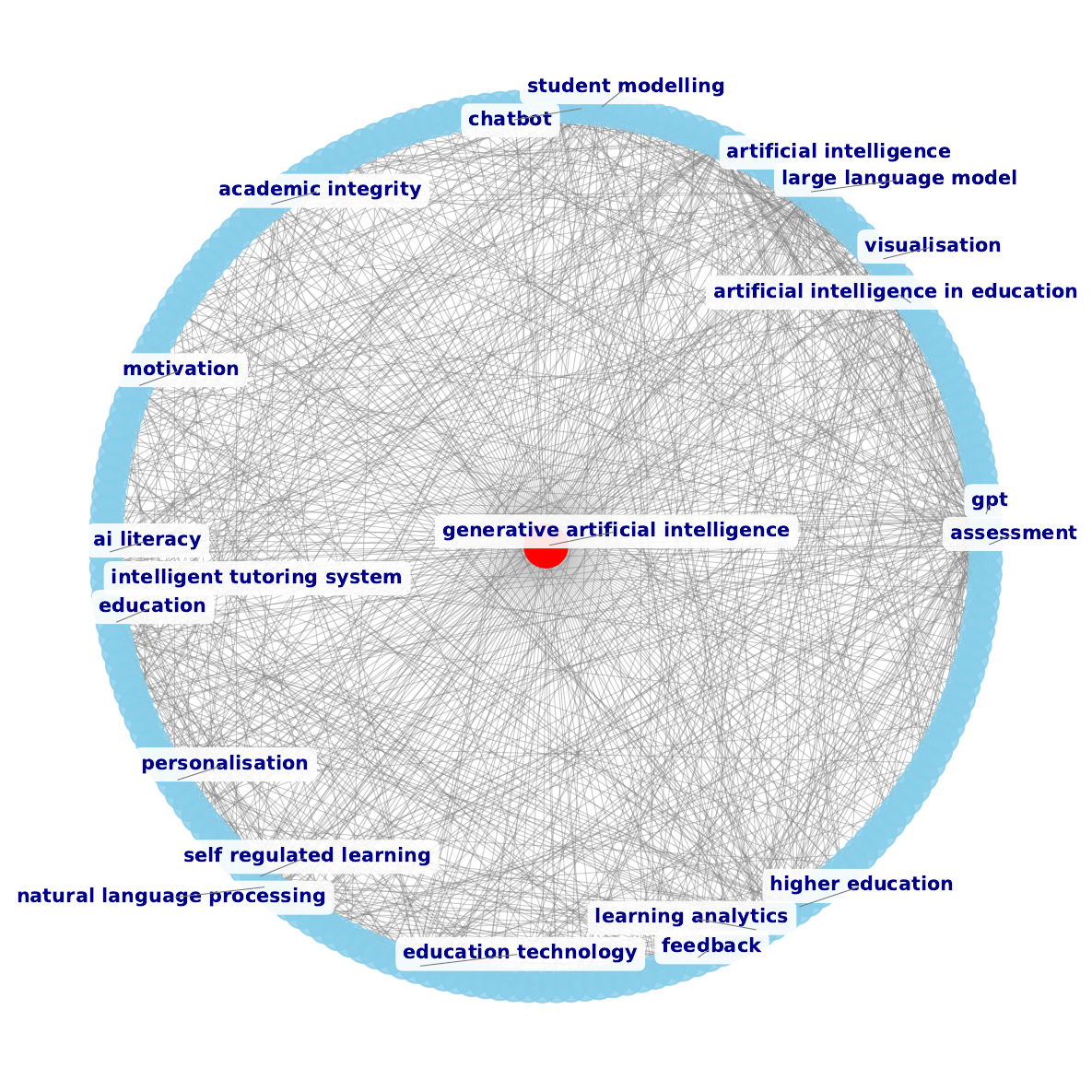}
    \caption{The ego network of the second emerging theme \textit{generative artificial intelligence}, with labels for the top 20 alters by degrees.
    }
    \label{fig:diagram}
\end{figure*}

In learning analytics—which involves a cyclical process of learners, data, metrics, and intervention (Clow, 2012)—GenAI enhances functionality by processing unstructured data, generating synthetic learning data, supporting multimodal interactions, and improving the interactivity and interpretability of analyses, thereby facilitating personalized and adaptive interventions (Yan et al., 2024). In assessment and feedback, existing studies have employed multimodal large models in GenAI (e.g., text, images, speech, video) to enhance interactivity and formative assessment efficiency (Lin et al., 2024; Cropley \& Marrone, 2022). Moreover, GenAI holds promise for empowering self-regulated learning by supporting students in goal setting and planning based on their behavioral data, monitoring and analyzing their learning behaviors through automated analysis of thought processes, and providing interactive feedback, supplementary information, and guidance for reflection through dialogue (Goslen et al., 2024; Kumar et al., 2024; Zhang et al., 2024b). Notably, ethical issues—especially academic integrity—are prominent in GenAI research. AI-assisted writing increases risks of plagiarism and undermines originality, highlighting the need for strict policies and education  (Li et al., 2024).\\

The third emerging topic is \textbf{Multimodal learning analytics (MMLA)} (see Figure 10), which involves collecting and integrating diverse data sources to comprehensively understand learning and its complex processes (Blikstein \& Worsley, 2016). MMLA has found applications in multiple domains such as computer programming, science games and online learning, with particular focus on collaborative learning (Fonteles et al, 2024; Ma et al., 2022). Unlike traditional analytics—restricted to log data or test scores—MMLA captures the nuanced group dynamics essential to knowledge co-construction. By integrating various forms of data, MMLA reconstructs dynamic interactions, individual roles, emotional states, conflicts, and rapport within groups (Chejara et al., 2023; Fahid et al., 2023; Wang et al., 2024c).  For example, Ma et al. (2022) employed linguistic, audio, and video modalities to jointly detect impasse states during students’ collaborative problem-solving processes. Wang et al. (2024c) analyzed collaborative patterns among students in maker activities using electroencephalography (EEG), eye-tracking, and system log data. Li et al. (2024d) designed the mBox learning platform, which leverages low-cost wearable badges and Internet of Things (IoT) technologies to enable the collection, synchronization, and analysis of multimodal data from collaborative learning groups. Moreover, the abundance of data can increase users’ cognitive load and has prompted research into more effective presentation of complex datasets through learning analytics dashboards. Yan et al. (2024e) introduced VizChat, a system capable of visualizing multiple sources of information, such as student locations and student-patient dialogues. By transparently explaining data analysis processes and providing personalized responses, VizChat offers robust support for educational decision-making. In particular, the selection of data types and analytic methods is crucial. Commonly employed data modalities include behavioral data, physiological signals, eye-tracking metrics, facial expressions, audio features, textual data and spatial data representing learners’ physical locations (Chejara et al., 2023; Feng et al., 2024; Fonteles et al., 2024; Lämsä et al., 2024; Ma et al., 2022; Schneider \& Sung, 2024). In multimodal learning analytics, it is meaningful to compare different modalities as well as to contrast unimodal with multimodal models (Emerson et al., 2023; Mangaroska et al., 2020; Rajarathinam et al., 2024). Empirical evidence indicates that multimodal data substantially improve learning outcome predictions and uncover behavioral patterns undetectable by unimodal approaches (Acosta et al., 2024; Fahid et al., 2023). \\

 \begin{figure*}
    \centering
    \includegraphics[width=0.7\textwidth]{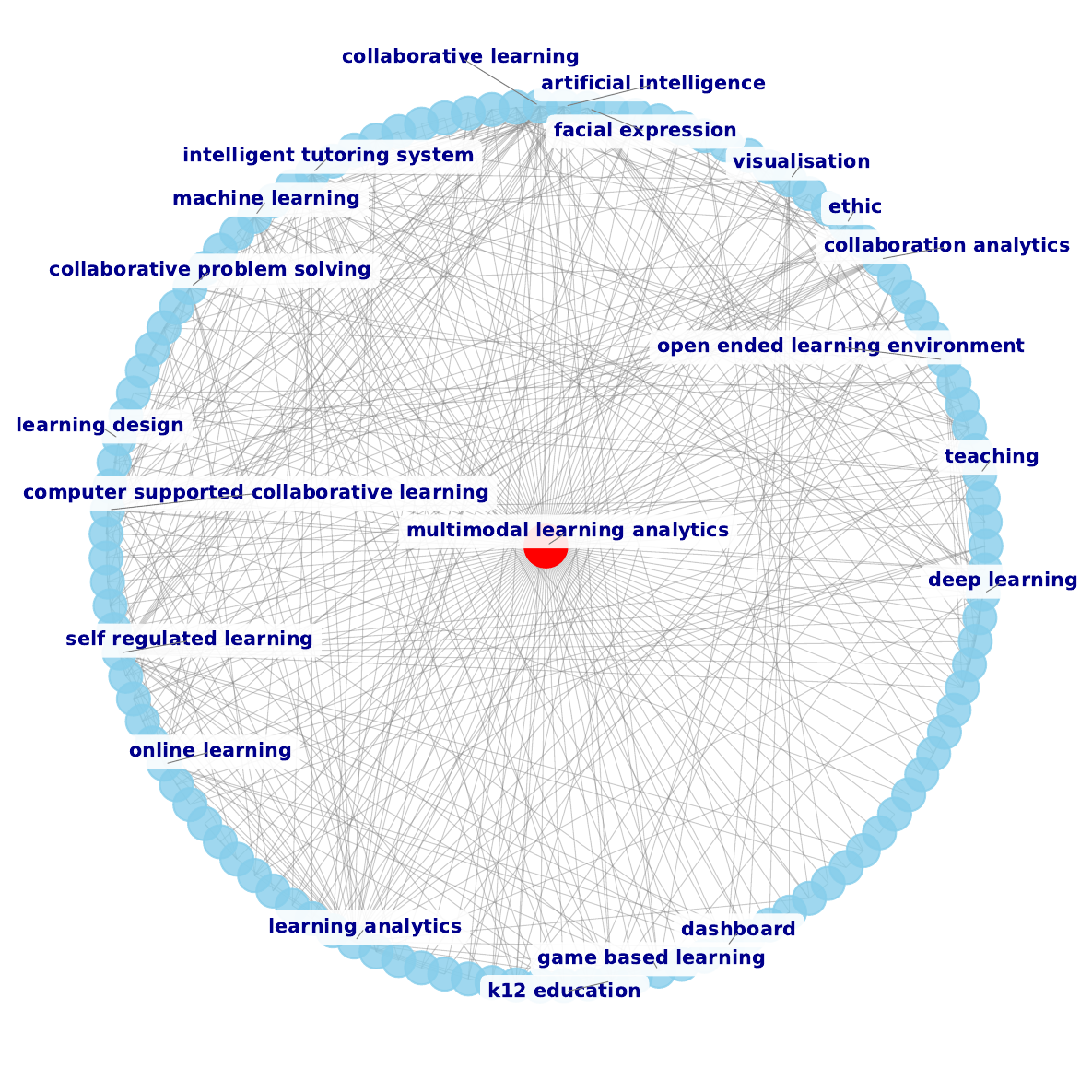}
    \caption{The ego network of the third emerging theme \textit{multimodal learning analytics}, with labels for the top 20 alters by degrees.
    }
    \label{fig:diagram}
\end{figure*}

 The fourth emerging topic in AIED field is \textbf{Human-AI collaboration} (see Figure 11), involving joint efforts between human intelligence and AI algorithms in knowledge work to reach informed results (Puranam, 2021). On the one hand, research highlights the importance of involving users to improve AI outcomes and correct or supplement AI judgments (Myint et al., 2024; Sun et al., 2024). Aslan et al. (2024) designed "Kid Space", a multimodal and immersive game-based learning system aimed at enhancing children's mathematical performance while promoting physical activity and social interaction. Since deep learning components have not yet achieved human-level understanding in dialogue systems, human experts monitor all learning content and student activities in the space and promptly intervene to correct any errors made by the system (Mosqueira-Rey et al., 2023). On the other hand, AI can simplify human actions and enhance work efficiency. According to Brusilovsky (2024), AI supports teacher and learner decisions through mechanisms such as ranking relevant content, providing annotations and explanations, and offering navigation guidance to optimize learning paths. Human-AI dialogue is an important form of achieving human-AI collaboration. Dialogue, primarily based on language models, enables students to engage in personalized interactions with conversational agents. By reflecting on and adjusting their study based on the machine's responses in  scenarios such as collaborative programming, students can enhance their learning outcomes and confidence (Kumar et al., 2024; Sankaranarayanan et al., 2021).  Human-AI collaborative learning presents new opportunities and challenges for learning analytics. Technologies such as GenAI may mimic certain learning processes and produce learning outcomes, making it challenging to define "the very essence of what it means to be a learner" (Yan et al., 2024d, p.103). Therefore, learning analytics should not only capture human data but also gather computational data to explore the collaborative processes effectively (Yan et al., 2024d). Beyond these technical dimensions, researchers also investigate human attitudes toward AI. AI's rapid advancement naturally generates anxiety, manifesting in two distinct responses: some avoid AI use altogether, while others are motivated to develop better collaboration skills with these technologies (Kaya et al., 2024). Xu et al. (2023) explored human-centered design and careful presentation of algorithm-based recommendations can reduce resistance from users—particularly domain experts—toward machine-generated decisions. Guggemos (2024) suggests our focus should shift from fearing AI replacement to leveraging AI to enhance everyday work.

\begin{figure*}
    \centering
    \includegraphics[width=0.7\textwidth]{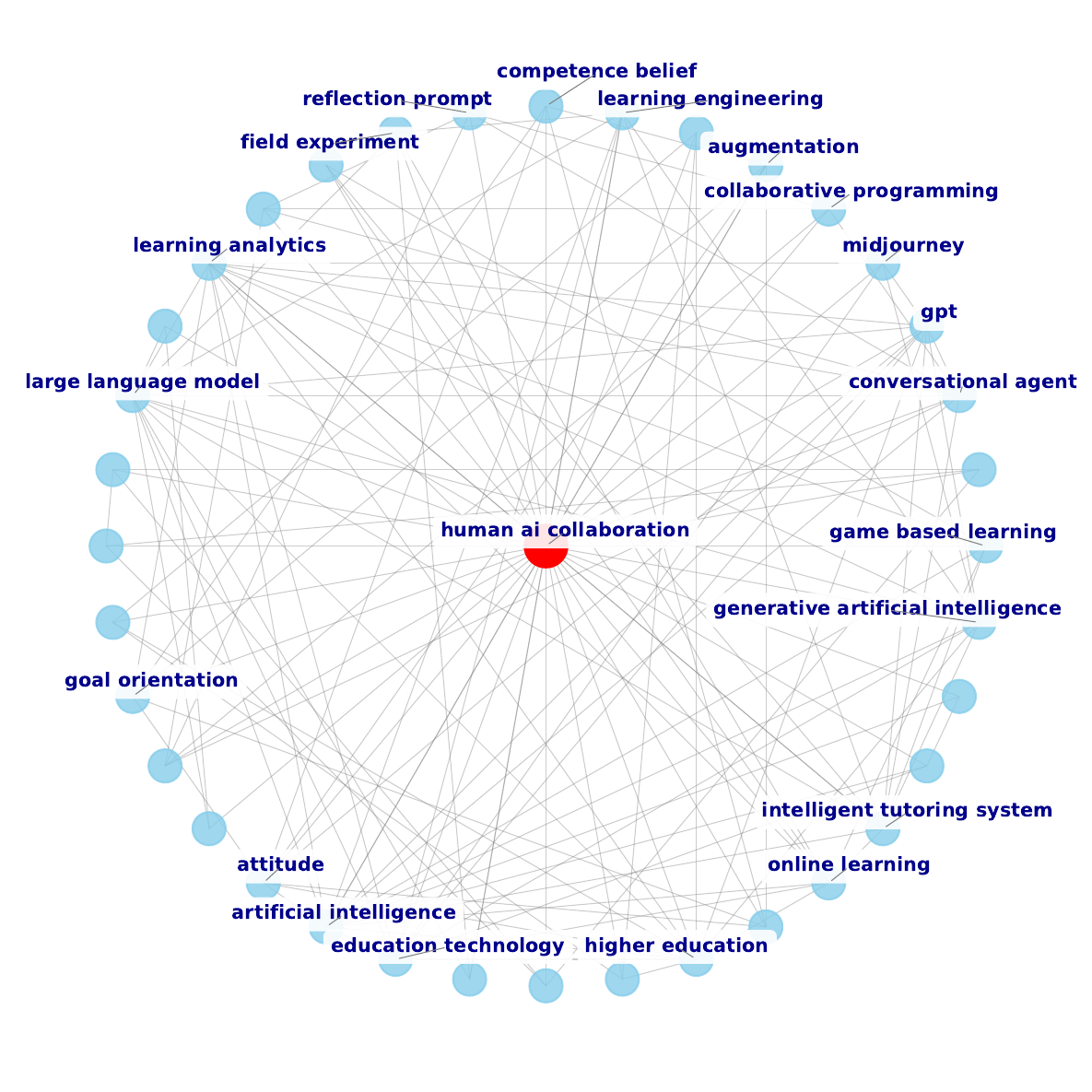}
    \caption{The ego network of the third emerging theme \textit{human AI collaboration}, with labels for the top 20 alters by degrees.
    }
    \label{fig:diagram}
\end{figure*}

\section{Discussion}

This study systematically analyzed the current focus and emerging topics in AIED field.  From a macro perspective, as showed by the results related to RQ1, the knowledge structure of AIED research exhibited a hierarchical structure with several core keywords as hubs linking a wide variety of research topics. At the meso level, we identified sustained themes in AIED, including natural language processing and intelligent tutoring systems, while also uncovering recent emerging knowledge clusters. At the micro level, we employed betweenness centrality rankings to highlight the pivotal roles of keywords in connecting distinct research themes and to identify emerging hotspots, including large language models, generative AI, multimodal learning analytics, and human-AI collaboration. \\

The identified sustained knowledge clusters represent the core concepts in AIED research, including natural language processing, intelligent tutoring systems, massive open online courses, lifelong learning, and self-regulated learning.  Natural language processing plays a pivotal role in facilitating interactive and personalized learning experiences through chatbots and intelligent tutoring systems. Deep learning-based NLP is becoming a core technology in Human-AI interaction and will continue to evolve towards more accurate sentiment analysis, greater interpretability, and deeper integration with other modalities, ultimately optimizing the interaction process between humans and intelligent agents (Ahmed et al., 2024a; Chen et al., 2024). Intelligent tutoring systems embody the longstanding pursuit of personalized and high-quality intelligent teachers. Although existing ITS continue to face challenges related to scalability and cognitive adaptability gaps (Liu et al., 2025), they played a significant role in education through robust data mining and human-computer collaboration (Lin et al., 2023). Massive open online courses provide access to quality learning resources for diverse populations, thereby promoting the democratization of education to some extent (Barger, 2020; Li, 2021). Consequently, exploring how to effectively deliver valuable online courses that can meet the lifelong learning needs in the rapidly changing social landscape will continue to be a focal point of AIED research. Self-regulated learning emphasizes individual responsibility for one’s own learning, allowing us to maintain autonomy in an era of artificial intelligence (Giannakos et al., 2024). SRL also extends to the group level, evolving into socially shared regulation of learning (SSRL), which highlights the importance of interactive planning and reflection within groups, thereby aligning with the needs of collaborative learning (Edwards et al., 2024). By interpreting multimodal data within the SRL process, we can more accurately capture the cognitive and metacognitive processes, understand the cyclical patterns of regulatory behaviors, and explore the interaction between cognitive and non-cognitive factors that are often difficult to observe, facilitating finer-grained analyses (Fan et al., 2022; Järvelä et al., 2019). Overall, these enduring themes collectively reflect the educational ideals and foundational methods essential for achieving educational objectives. These themes also intersect with various emerging topics, highlighting their inclusiveness and potential for development.\\

The emerging keywords reveal the research frontiers in the field, including large language models, multimodal learning analytics, generative artificial intelligence, and human-AI collaboration. Large language models, as one of the most impactful applications of generative artificial intelligence, along with other breakthroughs in GenAI's ability to comprehend and generate information across diverse data sources—including text, speech, and images—have significantly expanded the scope of intelligent education. Nonetheless, despite their strong automation performance, GenAI faces challenges related output accuracy, domain relevance, algorithmic bias and limited availability of high-quality datasets (Xing et al., 2025; Blanchard \& Mohammed, 2024). Future research should prioritize the construction of comprehensive, high-quality training datasets, the development of more flexible model control, enhancing AI interpretability, and effective strategies for multimodal integration in generative models (Gašević et al., 2023; Zhang et al., 2025).\\

The emergence of multimodal learning analytics reflects AIED research's ongoing exploration for more precise measurement and interpretation of complex learning environments. It addresses the need to collect student data in non-virtual learning environments (Ouhaichi et al., 2023), and can offer deeper insights into (meta)cognition, emotions, and behaviors (Giannakos et al., 2022). Meanwhile, challenges remain, notably the complexity of multimodal data that hinders understanding, the intrusiveness of sensors disrupting learning, and the diminished data quality in authentic classroom settings due to noise from multiple sources (Acosta et al., 2024; Mangaroska et al., 2020). Looking ahead, the advancement of MMLA will depend on developing methods for unobtrusive, high-quality data collection in real classrooms and on innovating analytic techniques capable of handling complex, noisy, and heterogeneous data streams. Researchers also need to design intuitive visualizations to enhance user understanding and usability (Yan et al., 2024b).\\

Human-AI collaboration emphasizes mutual learning, cooperation, and reinforcement between humans and machines (Järvelä et al., 2023a), where AI should possess the following characteristics: explainable AI, user-friendly AI, and responsible AI (Jiang et al., 2024). Despite the positive effects of Human-AI collaboration, various factors such as “algorithm appreciation” and “algorithm aversion” may occasionally render collaboration less effective than the performance of tasks executed solely by either AI or humans (Eisbach et al., 2023; Vaccaro et al., 2024). Furthermore, Wu et al. (2025) highlighted through experimental research that while collaboration between humans and GenAI can enhance immediate task performance, it may compromise the long-term psychological experience of human workers due to the effects of psychological deprivation. \\

Additionally, accountability in Human-AI collaboration has garnered attention, emphasizing the need to establish regulatory standards to ensure the responsible design and use of AI (Gal-Or, 2025).  In the future, the theoretical exploration of human-AI collaboration requires a deeper integration of multidisciplinary theories and multi-tiered approaches, while its application may focus on the diverse roles of AI and the capacity to address complex tasks (Jiang et al., 2024). Furthermore, through advancements in touch perception, dexterity, human-robot interaction, as well as GenAI and multimodel data, embodied intelligence is expected to develop further, enabling deeper interactions between students and computers in both virtual environments and the physical world (Meta, 2024; Nguyen et al., 2025). It is also important for human collaborators to enhance their AI literacy.\\

Whether it involves GenAI, multimodal learning analytics, or human-AI collaboration, many researchers have consistently referenced the concept of "human-centered AI", emphasizing making AI understandable and controllable by humans to ensure reliability and safety (AI under human control), or designing algorithms with humanistic considerations at the core (AI on the human condition) (Yang et al., 2021). Researchers explore ways to achieve human-centred learning analytics through stakeholder co-creation, empowering users, and enhancing the security, reliability, and trustworthiness of learning analytics systems (Alfredo et al., 2024; Shum et al., 2019). Järvelä et al. (2023) also pointed out that we need to build hybrid intelligent systems that enhance rather than replace human intelligence, based on our understanding of human learning and interaction processes. 

\section{Conclusion}

This study contributes an updated and systematic analysis of the AIED landscape, and by identifying the current research foci and emerging topics. The findings indicate that research in AIED exhibits both stability and considerable dynamism, driven by technological advancements and changing educational demands. Core areas such as intelligent tutoring systems, learning analytics, natural language processing, and MOOCs continue to play key roles in improving learning outcomes and optimizing learner experience. Furthermore, the rapid increase in research output and knowledge integration related to large language models, GPT, and GenAI signals a more significant influence on education than earlier technologies. GenAI facilitates flexible, anytime-anywhere learning experiences, enabling a critical shift toward more open, personalized, and accessible education. Accordingly, students, teachers, and educational administrators should prepare to embrace human-AI collaboration both psychologically and practically; that is, we need not only to educate ourselves about GenAI but also to become autonomous individuals equipped with critical thinking, self-regulation, and problem-solving skills (Yan, et al., 2024c).\\

Several limitations should be noted. First, this study primarily relied on keyword-based network analysis supported by metadata on publication date and venue, without incorporating other bibliometric attributes. Future research could enrich the analysis by including additional indicators such as author information and citation networks to better understand academic collaboration and knowledge dissemination pathways. Second, we employed a keyword co-occurrence network to comprehensively cover significant themes in recent years, depicting the overall development trends in AIED research through knowledge structures, knowledge clusters, and trending topics. However, this breadth often limits our ability to provide in-depth analyses of each identified theme or emerging topic. Future research could enable further exploration of these important themes, yielding richer insights.


\appendix
\section{Appendix}
\begin{table}[htbp]
\centering
\caption{Top ten highest in-group degree keywords for each knowledge cluster}
\label{tab:example}
\begin{tabular}{|p{3cm}|p{5cm}|p{3cm}|p{5cm}|} 
\hline
\textbf{Clusters} & \textbf{Top ten highest in-group degree keywords} & \textbf{Clusters} & \textbf{Top ten highest in-group degree keywords} \\ 
\hline
1.'natural language processing' & 'machine learning',
'large language model',
'deep learning',
'gpt',
'assessment',
'knowledge tracing',
'education',
'bert',
'neural network',
'item response theory'
 & 6.“intelligent tutoring system” &  'ontology',
 'student model',
 'web ontology language',
 'pyactr',
 'cognitive agent',
 'task ontology',
 'domain ontology',
 'teacher dashboard',
 'act r',
 'situation awareness'
 \\
\hline
2.‘learning
analytics'
 & 'self-regulated learning',
'collaborative learning',
'multimodal learning analytics',
'student modelling',
'education data mining',
'collaborative problem solving',
'feedback',
'clustering',
'motivation',
'computer science education'
 & 7. 'causal inference' & 'programming',
'user experience',
'hint',
'self explanation',
'scheduling',
'python',
'productive
'persistence',
'tutoring system',
'data driven method',
'worked example'
\\
\hline
3.'massive open online course' & 'online learning',
'higher education',
'ai literacy',
'learning management system',
'ethic',
'online education',
'privacy',
'personalisation',
'electroencephalograph',
'dashboard'
 & 8.'lifelong learning' & 'intelligent tutoring',
'adult learning',
'curriculum analytics',
'research method at scale',
'introductory programming course',
'facial emotion recognition',
'principal component analysis',
'social interaction',
'discussion board',
'problem decomposition'
 \\
\hline
4.“artificial intelligence” & 'generative artificial intelligence',
'artificial intelligence in education',
'student engagement',
'technology acceptance model',
'systematic review',
'computer assisted language learning',
'decision making',
'social medium',
'english language teaching',
'technology in education'
 & 9.'computer vision' & 'authoring tool',
'classroom analytics',
'augmented reality',
'mixed reality',
'adaptive learning technology',
'conversation analysis',
'design analytics',
'spatial analytics',
'stop detection',
'hyperparameters'
 \\
\hline
5.“engagement”  & 'eye tracking',
'classification',
'topic modelling',
'discussion forum',
'online discussion',
'question generation',
'critical thinking',
'early warning system',
'dialogue system',
'attention'
& 10.'data science application in education' & 'evaluation methodology',
'game',
'cooperative/collaborative learning',
'bias mitigation',
'human computer interface',
'post secondary education',
'distributed learning environment',
'improving classroom teaching',
'supervised machine learning',
'application in subject area'
\\
\hline
\end{tabular}
\end{table}

\clearpage
\onecolumngrid

\end{document}